\documentclass[floafix,preprint,showpacs,showkeys,preprintnumbers,nofootinbib,superscriptaddress]{revtex4-1}
\usepackage[utf8]{inputenc}
\usepackage[sort&compress]{natbib}
\usepackage{ulem}
\usepackage{bm}
\usepackage{times}
\usepackage{amssymb,amsbsy,amsmath,amsfonts}
\usepackage{graphicx}
\usepackage{epstopdf}
\usepackage{float}
\usepackage{color}
\usepackage{morefloats}
\usepackage{rotating}
\usepackage{srcltx}
\usepackage{slashed}
\usepackage{subfigure}
\usepackage{multirow}
\usepackage{verbatim}
\usepackage{hyperref}
\usepackage{overpic}
\usepackage{tabularx}

\begin{document}

\title{Can the nature of $a_0(980)$ be tested in the  $D_s^{+}\to \pi^{+}\pi^0 \eta$ decay?}

\author{Xi-Zhe Ling}
\affiliation{School of Physics, Beihang University, Beijing 100191, China}

\author{Ming-Zhu Liu}
\affiliation{School of Space and Environment, Beihang University, Beijing 100191, China}
\affiliation{School of Physics, Beihang University, Beijing 100191, China}

\author{Jun-Xu Lu}
\affiliation{School of Physics, Beihang University, Beijing 100191, China}

\author{Li-Sheng Geng}~\email{lisheng.geng@buaa.edu.cn}
\affiliation{School of Physics, Beihang University, Beijing 100191, China}
\affiliation{Beijing Key Laboratory of Advanced Nuclear Materials and Physics,
Beihang University, Beijing 100191, China}
\affiliation{School of Physics and Microelectronics, Zhengzhou University, Zhengzhou, Henan 450001, China}

\author{Ju-Jun Xie}~\email{xiejujun@impcas.ac.cn}
\affiliation{Institute of Modern Physics, Chinese Academy of Sciences, Lanzhou 730000, China}
\affiliation{School of Nuclear Science and Technology, University of Chinese Academy of Sciences, Beijing 101408, China}
\affiliation{School of Physics and Microelectronics, Zhengzhou University, Zhengzhou, Henan 450001, China}

\date{\today}
\begin{abstract}

From the amplitude analysis of the $D^+_s \to \pi^+ \pi^0 \eta$ decay, the BESIII Collaboration  firstly observed the $D^+_s \to a_0(980)^+\pi^0$ and  $D^+_s \to a_0(980)^0\pi^+$ decay modes, which are expected to occur through the pure $W$-annihilation processes. The measured branching fraction $\mathcal{B}[D_{s}^{+}\to a_{0}(980)^{+(0)}\pi^{0(+)},a_{0}(980)^{+(0)}\to \pi^{+(0)}\eta]$ is, however, found to be larger than those of known $W$-annihilation decays by one order of magnitude. This apparent contradiction can be reconciled if the two decays are induced by internal $W$-conversion or external $W$-emission mechanisms instead of $W$-annihilation mechanism. In this work, we propose that the $D^+_s$ decay proceeds via both the external and internal $W$-emission instead of $W$-annihilation mechanisms. In such a scenario, we perform a study of the $D^+_s \to \pi^+\pi^0\eta$ decay by taking into account the contributions from the tree diagram $D^+_s \to \rho^+ \eta \to \pi^+ \pi^0 \eta$ and the intermediate $\rho^+ \eta$ and $K^*\bar{K}/K\bar{K}^*$ triangle diagrams. The intermediate $a_0(980)$ state can be dynamically generated from the final state interactions of coupled $K \bar{K}$ and $\pi \eta$ channels, and it is shown that the experimental data can be  described fairly well, which supports the interpretation of $a_0(980)$ as a molecular state.

\end{abstract}


\maketitle

\section{Introduction}

The charmed meson weak decays into light mesons provide a very good channel to study meson-meson interactions at low energies and the nature of the low-lying scalar mesons~\cite{Klempt:2008ux,Bhattacharya:2010id,Liang:2016hmr,Debastiani:2016ayp}. We refer to Ref.~\cite{Oset:2016lyh} for a review about the study on the interactions of light hadrons from the weak decays of the $D$ mesons. Very recently, in Ref.~\cite{Ablikim:2019pit}, the BESIII Collaboration firstly reported on the observation of the decay modes of $D_{s}^{+} \to a_{0}(980)^{+}\pi^{0}$ and $D_{s}^{+}\to a_{0}(980)^{0}\pi^{+}$ in the amplitude analysis of the  $D_{s}^{+} \to \pi^{+}\pi^{0}\eta$ decay. The $D_{s}^{+} \to a_{0}(980)^{+(0)}\pi^{0(+)}$ decays are claimed as  $W$-annihilation dominant processes. However, the measured absolute branching fraction of $\mathcal{B}[D_{s}^{+} \to a_{0}(980)^{+(0)}\pi^{0(+)}$, $a_{0}(980)^{+(0)} \to \pi^{+(0)}\eta] = (1.46\pm 0.15_\mathrm{sta.}\pm0.23_\mathrm{sys.})\%$ is found to be larger than those of normal $W$-annihilation processes by at least one order of magnitude. In Ref.~\cite{Molina:2019udw} it was proposed that the $D^+_s \to \pi^+ \pi^0 \eta$ decay actually occurs via an internal $W$-conversion process, where $\pi K \bar{K}$ was produced at the first step, and then the $\pi \eta$ state is generated from final state interaction of $K\bar{K}$ in $S$-wave and isospin $I=1$, and the $a_{0}(980)$ resonance is dynamically generated via $\bar{K}K$ and $\pi\eta$ coupled channel interactions as described in the unitary chiral theory~\cite{Oller:1997ti,Oller:2000fj}. The main purpose of Ref.~\cite{Molina:2019udw} is to get the $a_0(980)$ signal, and thus, only the experimental data with the cut $M_{\pi^+\pi^-} > 1$ GeV were studied, for which the  tree diagram $D^+_s \to \rho^+ \eta \to \pi^+\pi^0 \eta$ does not contribute. On the other hand, in Ref.~\cite{Hsiao:2019ait} it was shown that the experimental measurements can also be described as an external $W$-emission process $D^+_s \to \rho^+ \eta$. Then the $\rho$ meson decays into a pair of $\pi\pi$ and the $\pi\eta$ pair fuses into $a_0(980)$, which subsequently decays into $\pi\eta$ again. As both the $W$-internal conversion and $W$-external emission processes are believed to be larger than the $W$-annihilation process, the puzzle seems to be resolved, though the two theoretical studies seem to give conflicting results regarding the responsible weak decay mechanism.

In the present work, we revisit this issue and argue that the two theoretical works are not necessarily  contradicting with each other. As a matter of fact, the triangle mechanism of Ref.~\cite{Hsiao:2019ait} may offer a way to estimate the unknown weak decay coupling  between $D_{s}^{+}$ and $\bar{K}K^*(\bar{K}^*K)$, i.e., the first vertex of the triangle diagrams for $D^+_s\to K^+\bar{K}^{*0}/K^{*+}\bar{K}^0\to\pi^+\pi^0\eta$.

\begin{figure}[htpb]
  \centering
  \includegraphics[scale=0.18]{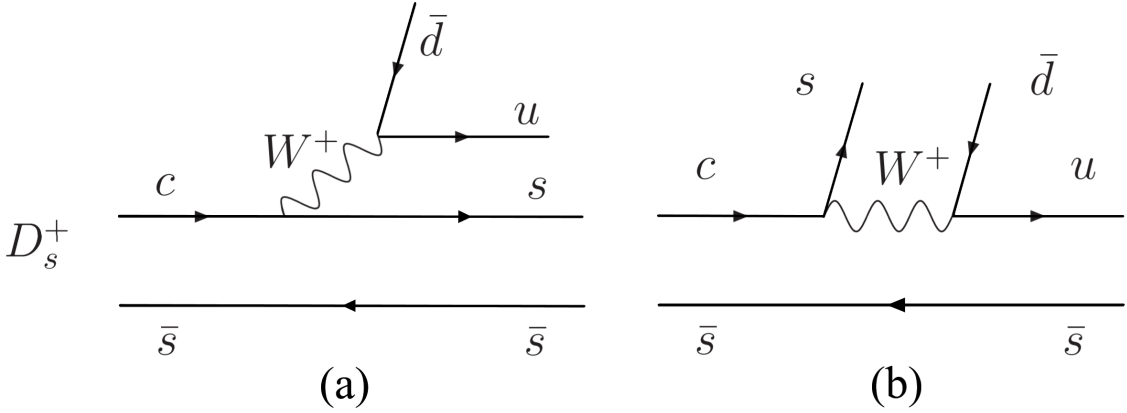}
  \caption{(a) External $W$-emission mechanism for $D^+_s\to \rho^+\eta$ and (b) internal $W$-conversion mechanisms for $D^+_s\to K^+\bar{K}^{*0}/K^{*+}\bar{K}^0$.}\label{quark}
\end{figure}

At the quark level, the decay of $D_{s}^{+}\to \rho^{+}\eta$ proceeds through external $W$-emission $c\to \bar{s} W^{+}$ as shown in Fig~\ref{quark}(a).  According to the review of the Particle Data Group~\cite{Zyla:2020zbs}, the absolute branching fractions of the decay modes $D_{s}^{+}\to K^{\ast +}\bar{K}^{0}$ and $D_s^+\to K^+\bar{K}^{*0}$ are $(5.4 \pm 1.2)\%$ and $(5.16 \pm 0.16)\%$ respectively, which are comparable to the absolute branching fraction of $D_{s}^{+}\to \rho^{+}\eta$ that is $(8.9 \pm 0.8)\%$. As a result, if $D^+_s\to \rho^+\eta$ can contribute to the $D^+_s\to a_0^{+(0)}\pi^{0(+)}$ [$a_0 \equiv a_0(980)$] process via the triangle diagrams as in Ref.~\cite{Hsiao:2019ait}, the $D^+_s\to K^+ \bar{K}^{*0}/\bar{K}^0K^{*+}$ processes shown in Fig~\ref{quark}(b) can also contribute via the triangle diagrams, where the $K^*$ decays into $K\pi$ and $K\bar{K}$ produce the $\pi \eta$ pair through final state interactions, from which the $a_0(980)$ resonance can be produced. The latter has not been considered either in Ref.~\cite{Molina:2019udw} or  in Ref.~\cite{Hsiao:2019ait}.~\footnote{As we will show later, the anomalously large $a_0\pi\eta'$ coupling adopted in Ref.~\cite{Hsiao:2019ait} helps to increase the branching fraction to meet the experimental number.} As a result, in this work we will consider both $D_{s}^{+} \to \rho^{+}\eta$  and $D_s\to K^+\bar{K}^{*0}/K^{*+}\bar{K}^0$ induced triangle diagrams, which lead to $\pi^+a_0^0$ and $\pi^0 a_0^+$ final states.

Compared to Ref.~\cite{Hsiao:2019ait}, we make a further improvement. It is well known that the $a_0(980)$ state does not behave like a normal Breit-Wigner resonance, because of the closeness of the $K\bar{K}$ threshold. It can be dynamically generated as a molecular state from the $K\bar{K}$ and $\pi\eta$ coupled channel interactions in the chiral unitary approach~\cite{Oller:1997ti,Oller:1997ng,Kaiser:1998fi,Locher:1997gr,Nieves:1999bx}. In this work, we then investigate whether with the chiral unitary amplitudes one can describe the BESIII data~\cite{Ablikim:2019pit}.

The article is organized as follows. In Sec.~\ref{sec:Interactions}, we lay out the theoretical formalism. In Sec.~\ref{sec:Results} we show our theoretical results and discussions are also given comparing with the experimental data from Ref.~\cite{Ablikim:2019pit}. We summarize in Sec. \ref{sec:Summary}.

\section{Theoretical FORMALISM}\label{sec:Interactions}

\begin{figure}[htbp]
  \centering
  \includegraphics[scale=0.2]{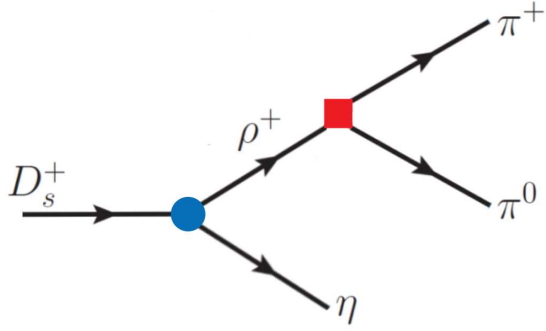}\\
  \caption{Tree-level diagram for $D_s^{+}\to \eta\rho^{+} \to \eta \pi^0\pi^{+}$.}\label{tree}
\end{figure}

\begin{figure}[htpb]
  \centering
  \includegraphics[scale=0.3]{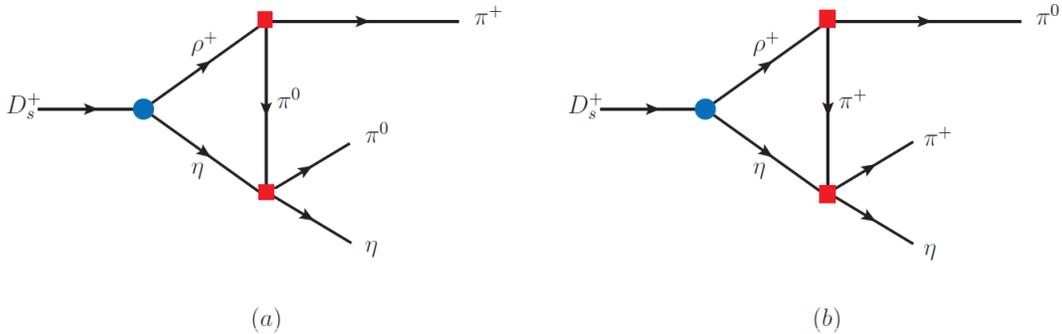}\\
  \caption{Triangle rescattering diagrams for $D_s^{+} \to (\rho^+\eta \to)\pi^+\pi^0\eta$.}\label{triangle}
  \end{figure}

\begin{figure}[htpb]
  \centering
   \includegraphics[scale=0.3]{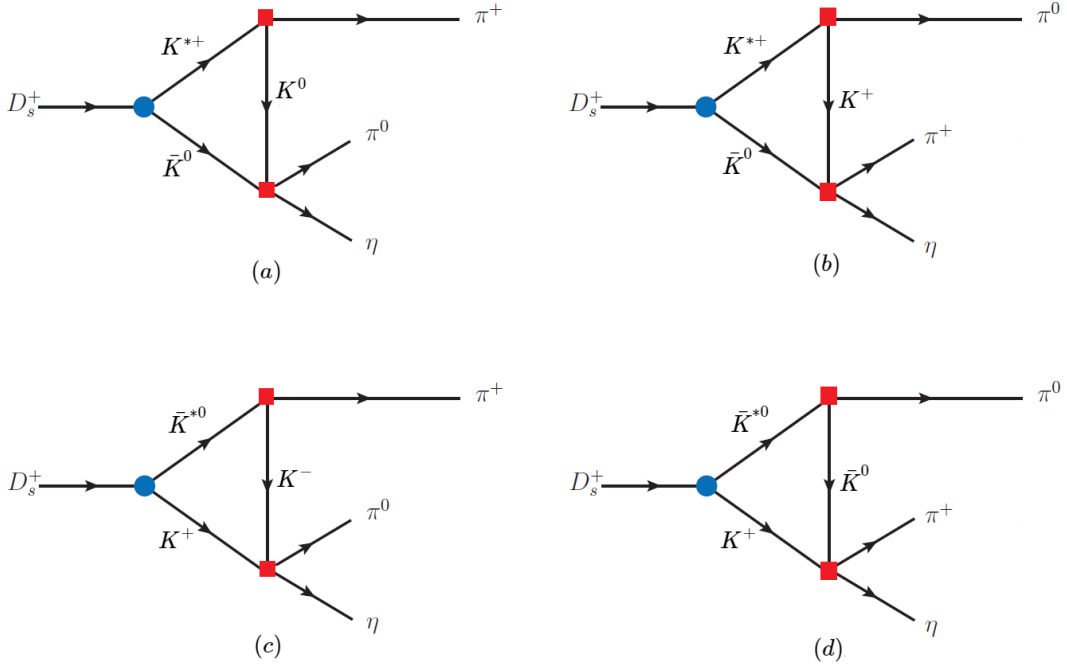}\\
  \caption{Triangle rescattering diagrams for $D_s^{+}\to (K^{*+}\bar{K}^0 \to)\pi^{+}\pi^0\eta$ and $D_s^{+}\to (K^{+}\bar{K}^{*0} \to)\pi^{+}\pi^0\eta$ .}\label{triangle KK}
\end{figure}

To calculate the decay width of $D_s^{+}\to \eta\pi^0\pi^{+}$, we  consider the contribution  from both the tree-level diagram of Fig.~\ref{tree} and the triangle diagrams of Fig.~\ref{triangle} and Fig.~\ref{triangle KK}. In the process described by the tree-level diagram,  $D_{s}^{+}$ first decays into $\rho^{+}$ and $\eta$, then $\rho^{+}$ decays to $\pi^{+}\pi^{0}$ as shown in Fig~\ref{tree}. As pointed out in Ref.~\cite{Hsiao:2019ait}, the $\pi^+$/$\pi^0$ meson can interact with the $\eta$ meson to form $a_0(980)^{+}/a_0(980)^0$ via the triangle diagrams shown in Fig.~\ref{triangle}. On the other hand, if the processes depicted in Fig.~\ref{triangle} can occur, those depicted in Fig.~\ref{triangle KK} can also occur, because 1) the branching fractions of $D_s$ decaying into $K^*\bar{K}/K\bar{K}^*$ are comparable to that of $D_s$ decaying into $\rho\eta$ and 2) $K\bar{K}$ is a dominant channel to which the $a_0(980)$ state couples. As a result, their contributions cannot be neglected. In this work, therefore, we also consider the contribution from $D_{s}^{+}\to K^{\ast+}\bar{K}^{0}/K^+\bar{K}^{*0}$, which can proceed through the triangle mechanisms  shown in Fig~\ref{triangle KK}.

The follwing effective Hamiltonian is needed to describe the $D_{s}^{+}\to \rho^{+}\eta$ and $ D_{s}^{+} \to K^{\ast+}\bar{K}^{0}/K^+\bar{K}^{*0}$ processes,
\begin{equation}
\mathcal{H}_{eff}=\frac{G_F}{\sqrt{2}}V_{cs}V_{ud}[c_1^{eff}(\bar{u}d)(\bar{s}c)+c_2^{eff}(\bar{s}d)(\bar{u}c)],
\end{equation}
where $G_F = 1.166 \times 10^{-5} {\rm GeV}^{-2}$ is the Fermi constant, $V_{cs}$ and $V_{ud}$ are the CKM matrix elements, $c_{1,2}^{eff}$ are the effective Wilson coefficients, and $\bar{q_1}q_2$ stand for vector and axial vector currents, $\bar{q_1}\gamma_\mu(1-\gamma_5)q_2$~\cite{Buras:1998raa,Geng:2017esc,Geng:2018upx,Geng:2020ofy,Han:2021azw}. The amplitude of $D_s^{+}\to \rho^{+} \eta$ and $D_{s}^{+} \to K^{\ast+}\bar{K}^{0}/K^+\bar{K}^{*0}$ can be written as the products of two hadronic matrix elements~\cite{Ali:1998eb,Li:2013xsa}
\begin{eqnarray}\label{Ds-KK}
\mathcal{A}\left(D_{s}^{+} \to \eta \rho^{+}\right)&=&\frac{G_{F}}{\sqrt{2}} V_{c s} V_{u d} a_{1}\left\langle\rho^{+}|(\bar{u} d)| 0\right\rangle\left\langle\eta|(\bar{s} c)| D_{s}^{+}\right\rangle\\
\mathcal{A}\left(D_s^{+}\to K^{*+}\bar{K}^0\right)&=&\frac{G_{F}}{\sqrt{2}} V_{c s} V_{u d}  a_{2}\left\langle \bar{K}^0|(\bar{s} d)| 0\right\rangle\left\langle K^{*+}|(\bar{u} c)| D_{s}^{+}\right\rangle, \\
\mathcal{A}\left(D_s^{+}\to \bar{K}^{*0}K^+\right)&=&\frac{G_{F}}{\sqrt{2}} V_{c s} V_{u d}  a_{2}' \left\langle \bar{K}^{*0}|(\bar{s} d)| 0\right\rangle\left\langle K^{+}|(\bar{u} c)| D_{s}^{+}\right\rangle.
\end{eqnarray}
where $a_{1}=c_{1}^{e f f}+c_{2}^{e f f} / N_{c}$ and $a_{2}=c_{1}^{e f f}/ N_{c}+c_{2}^{e f f} $ with $N_c$ the number of colors. It should be noted that $a_1$ can be calculated in the naive factorization approach, but $a_2$ and $a_2'$ cannot be easily obtained within the factorization approach~\cite{Beneke:2003zv,Ali:1998gb}. In the present work we will fix all of them by fitting directly to data.

 The current matrix elements between a pseudoscalar meson or vector meson and the vacuum have the following form:
\begin{eqnarray}
\left\langle\rho^{+}|(\bar{u} d)| 0\right\rangle =  m_{\rho} f_{\rho} \epsilon_{\mu}^{*} , ~~
\left\langle \bar{K}^0|(\bar{s} d)| 0\right\rangle = - f_{\bar{K}^0} p_{\bar{K}^0}^{\mu},  ~~
\left\langle \bar{K}^{*0}|(\bar{s} d)| 0\right\rangle = m_{\bar{K}^{*0}}f_{\bar{K}^{*0}}\epsilon_\mu^*,
\end{eqnarray}
where $f_{K}$, $f_{K^{\ast}}$, and $f_{\rho}$ are the decay constants for $K$, $K^{\ast}$, and $\rho$ mesons, respectively, and $\epsilon_\mu^*$ is the polarization vector of $\rho$ or $K^*$ meson. In this work, we take $f_\rho = 210$ MeV, $f_K =158$ MeV, $f_{K^*} = 214$ MeV as in Refs.~\cite{Ali:1998eb,Wang:2008rk,Ali:2007ff}.

The hadronic matrix elements can be written in terms of form factors as follows~\cite{Soni:2018adu}
\begin{eqnarray}
\left\langle\eta|(\bar{s} c)| D_{s}^{+}\right\rangle&=&\left(p_{D_{s}}+p_{\eta}\right)^{\mu} F_{+}(q^2)+q^{\mu} F_{-}(q^2), \\
\left\langle K^{*+}|(\bar{u} c)| D_{s}^{+}\right\rangle&=&\frac{\epsilon_{\alpha}^{*}}{m_{D_s}+m_{K^{*+}}}\left[-g^{\mu \alpha} P \cdot q' A_{0}\left(q'^{2}\right)+P^{\mu} P^{\alpha} A_{+}\left(q'^{2}\right)\right.\\
&&\left.+q'^{\mu} P^{\alpha} A_{-}\left(q'^{2}\right)+i \varepsilon^{\mu \alpha \beta \gamma}P_\beta q'_\gamma V\left(q'^{2}\right)\right], \\
\left\langle K^{+}|(\bar{u} c)| D_{s}^{+}\right\rangle &=&\left(p_{D_{s}}+p_{K^+}\right)^{\mu} F_{+}(q''^2)+q''_{\mu} F_{-}(q''^2),
\end{eqnarray}
where $q^{\mu(\prime,\prime\prime)}$ represent the momentum of $\rho$, $K$, and $K^*$ mesons, respectively, and $P^\mu = (p_{D_s} + p_{K^*})^{\mu}$. The form factors of $F_{\pm}(t)$, $A_{0}(t)$, $A_{+}(t)$, $A_{-}(t)$, and $V(t)$ with $t \equiv q^{(\prime,\prime\prime)2}$ can be parameterized as~\cite{Soni:2018adu}
\begin{equation}
X(t)=\frac{X(0)}{1-a\left(t / m_{D_{s}}^{2}\right)+b\left(t^{2} / m_{D_{s}}^{4}\right)}.
\end{equation}

In this work, we take these form factors $F_{\pm}$, $A_{0,\pm}$ and $V$ from Ref.~\cite{Soni:2018adu}: $(F_+(0), a, b)^{D_s \to \eta} = (0.78, 0.69, 0.002)$, $(F_+(0), a, b)^{D_s \to K} = (0.60, 1.05, 0.18)$, $(A_+(0), a, b)^{D_s \to K^*} = (0.57, 1.13, 0.21)$, $(A_0(0), a, b)^{D_s \to K^*} = (1.53, 0.61, -0.11)$, and $(A_-(0), a, b)^{D_s \to K^*} = (- 0.82, 1.32, 0.34)$. Note that the terms containing $V(q^{\prime 2})$ and $F_{-}(q^2(q^{\prime\prime 2}))$ do not contribute to the processes we study here.

For the strong decays $\rho^{+}\to\pi^{+}\pi^0$ and $K^{*+}\to K\pi$ ($\bar{K}^{*0}\to \bar{K}\pi$), the amplitudes are
\begin{align}
\mathcal{A}\left(\rho^{+} \to \pi^{+} \pi^{0}\right) = g_{\rho \pi \pi} \epsilon \cdot\left(p_{\pi^{+}}-p_{\pi^{0}}\right), ~~~
 \mathcal{A}\left(K^* \to K\pi\right) = g_{K^{*} K \pi} \epsilon \cdot\left(p_{\pi}-p_{K}\right),
\end{align}
where $g_{\rho\pi\pi}$ and $g_{K^{*} K \pi} $ denote the $\rho$ coupling to $\pi\pi$ and the $K^{*}$ coupling to $K\pi$. With the masses of these particles and the partial decay widths of $\rho \to \pi\pi$ and $K^* \to K \pi$ quoted in the PDG~\cite{Zyla:2020zbs}, we obtain $g_{\rho\pi\pi} = 6.0$, $g_{K^{*+} K \pi}=3.26$, and $g_{K^{*0} K \pi}=3.12$. On the other hand, the couplings~\footnote{It should be stressed that the partial decay widths determine only the absolute value of the corresponding coupling constants, but not their phases. In this work, we assume that they are real and positive, which seems to be a reasonable choice given the reasonable description of the experimental data as shown below.} $a_1=0.96$, $a_2=1.50$ and $a_2'=1.51$ are determined by fitting them to the experimental branching fractions $\mathcal{B}(D_s^+\to \rho^+\eta) = (8.9\pm0.8)\%$, $\mathcal{B}(D_s^+\to K^{*+}\bar{K}^0) = (5.4\pm 1.2)\% $ and $\mathcal{B}(D_s^+\to K^+\bar{K}^{*0}, \bar{K}^{*0}\to K^-\pi^+) = (2.58\pm 0.08)\%$ quoted in the PDG~\cite{Zyla:2020zbs}.

Putting all the pieces together, we obtain the decay amplitude of the $D^+_s \to \pi^+ \pi^0 \eta$ from the tree-level diagram of Fig.~\ref{tree} as
\begin{equation}
\mathcal{A}^{\rm tree}_1=i~\frac{\mathcal{A}(D_s^+\to\eta\rho^+)\mathcal{A}\left(\rho^{+} \to \pi^{+} \pi^{0}\right)}{m_{13}^2-m_{\rho^+}^2+im_{\rho^+}\Gamma_{\rho^+}},
\end{equation}
with $m_{13}^2=(p_{\pi^0}+p_{\pi^{+}})^2$ the invariant mass squared of the $\pi^+ \pi^0$ system.

Next, we can write the total decay amplitude of $D^+_s \to \pi^+ \pi^0 \eta$ for those triangle diagrams shown in Fig.~\ref{triangle},
\begin{align}\label{triangleA}
\mathcal{A}_{2}^{\rho \eta} & = \mathcal{A}_{a}^{\rho \eta}+\mathcal{A}_{b}^{\rho \eta}, \\
\mathcal{A}_{a}^{\rho \eta} & = t_{\pi\eta\to\pi\eta}(m_{12}^2) \int \frac{d^{4} q_{3}}{(2 \pi)^{4}} \frac{{\rm i}\mathcal{A}(D_s^+\to\eta\rho^+)\mathcal{A}\left(\rho^{+} \to \pi^{+} \pi^{0}\right)}{\left(q_{1}^{2}-m_{\rho^{+}}^{2}+i m_{\rho^{+}}\Gamma_{\rho^{+}}\right)\left(q_{2}^{2}-m_{\eta}^{2}+i \epsilon\right)\left(q_{3}^{2}-m_{\pi^{0}}^{2}\right)},  \\
\mathcal{A}_{b}^{\rho \eta} & =  t_{\pi\eta\to\pi\eta}(m_{23}^2)  \int \frac{d^{4} q_{3}}{(2 \pi)^{4}} \frac{{\rm i}\mathcal{A}(D_s^+\to\eta\rho^+)\mathcal{A}\left(\rho^{+} \to \pi^{0} \pi^{+}\right)}{\left(q_{1}^{2}-m_{\rho^{+}}^{2}+i m_{\rho^{+}}\Gamma_{\rho^{+}}\right)\left(q_{2}^{2}-m_{\eta}^{2}+i \epsilon\right)\left(q_{3}^{2}-m_{\pi^{+}}^{2}\right)},
\end{align}
where $m_{12}^{2}=\left(p_{\pi^0}+p_{\eta}\right)^{2}$, $m_{23}^{2}=\left(p_{\pi^{+}}+p_{\eta}\right)^{2}$, and the momenta $(q_1, q_2, q_3)$ are those of  $(\rho^+, \eta, \pi^{0(+)})$, respectively. The $t_{\pi\eta\to\pi\eta}$ stands for the two-body $\pi \eta \to \pi \eta$ scattering amplitude, which depends on the invariant mass of the $\pi \eta$ system.

The decay amplitudes of $D_{s}^{+} \to \pi^+ \pi^0 \eta $ via triangle diagrams shown in Fig.~\ref{triangle KK} are written as
\begin{eqnarray}\label{KK}
\mathcal{A}_{3}^{K^*K} &=& \mathcal{A}_{a}^{K^*K}+\mathcal{A}_{b}^{K^*K}+\mathcal{A}_{c}^{K^*K}+\mathcal{A}_{d}^{K^*K}, \\
\mathcal{A}_{a}^{K^*K} &  = & \frac{i}{\sqrt{2}} \int \frac{d^{4} q_{3}}{(2 \pi)^{4}} \frac{ \mathcal{A}\left(D_s^{+}\to K^{*+}\bar{K}^0\right) \mathcal{A}\left(K^{*+}\to \pi^{+}K^{0}\right)t_{K\bar{K}\to\pi\eta}(m_{12}^2) }{\left(q_{1}^{2}-m_{K^{*+}}^{2}+im_{K^{*+}}\Gamma_{K^{*+}}\right)\left(q_{2}^{2}-m_{K^0}^{2}+i \epsilon\right)\left(q_{3}^{2}-m_{K^{0}}^{2}\right)},  \\
\mathcal{A}_{b}^{K^*K} & = &  -i
 \int \frac{d^{4} q_{3}}{(2 \pi)^{4}} \frac{ \mathcal{A}\left(D_s^{+}\to K^{*+}\bar{K}^0\right) \mathcal{A}\left(K^{*+}\to \pi^{0}K^{+}\right) t_{K\bar{K}\to\pi\eta}(m_{23}^2) }{\left(q_{1}^{2}-m_{K^{*+}}^{2}+im_{K^{*+}}\Gamma_{K^{*+}}\right)\left(q_{2}^{2}-m_{K^0}^{2}+i \epsilon\right)\left(q_{3}^{2}-m_{K^{+}}^{2}\right)}, \\
\mathcal{A}_{c}^{K^*K} & = & -\frac{i}{\sqrt{2}}   \int \frac{d^{4} q_{3}}{(2 \pi)^{4}} \frac{\mathcal{A}\left(D_s^{+}\to \bar{K}^{*0}K^+\right) \mathcal{A}\left(\bar{K}^{*0}\to \pi^{+}K^{-}\right)t_{K\bar{K}\to\pi\eta}(m_{12}^2) }{\left(q_{1}^{2}-m_{K^{*0}}^{2}+im_{K^{*0}}\Gamma_{K^{*0}}\right)\left(q_{2}^{2}-m_{K^+}^{2}+i \epsilon\right)\left(q_{3}^{2}-m_{K^{-}}^{2}\right)}, \\
\mathcal{A}_{d}^{K^*K} &  = &  -i  \int \frac{d^{4} q_{3}}{(2 \pi)^{4}} \frac{\mathcal{A}\left(D_s^{+}\to \bar{K}^{*0}K^+\right) \mathcal{A}\left(\bar{K}^{*0}\to \pi^{0}\bar{K}^{0}\right)t_{K\bar{K}\to\pi\eta}(m_{23}^2) }{\left(q_{1}^{2}-m_{K^{*0}}^{2}+im_{K^{*0}}\Gamma_{K^{*0}}\right)\left(q_{2}^{2}-m_{K^+}^{2}+i \epsilon\right)\left(q_{3}^{2}-m_{K^{0}}^{2}\right)},
\end{eqnarray}
with momenta $(q_1, q_2, q_3)$ for $(K^*(\bar{K}^*), \bar{K}(K), K(\bar{K}))$, respectively. It is worth mentioning that one needs to include the isospin factor $-\sqrt{\frac{1}{3}}$ and $\sqrt{\frac{1}{3}}$ for Figs.~\ref{triangle KK}~(a) and (c) and Figs.~\ref{triangle KK}~(b) and (d), respectively. The $t_{K\bar{K}\to\pi\eta}$ stands for the two-body $K \bar{K} \to \pi \eta$ scattering amplitude, which depends on the invariant mass of the $\pi \eta$ system. It should be noted that in the present work, for the $\rho$ and $K^*$ vector meson propagators, we take
\begin{eqnarray}
G^{\mu\nu}_V(q^2_V) = \frac{i(-g^{\mu\nu} + q^{\mu}_{V} q^{\nu}_{V}/q^2_{V})}{q_{V}^2-m_{V}^2+im_{V}\Gamma_{V}},
\end{eqnarray}
where $m_V$ and $\Gamma_V$ are the mass and width of the vector mesons.

The triangle loop integrals in these above amplitudes are ultraviolet divergent, in general one needs to include phenomenological form factors to prevent ultraviolet divergence, as shown in Refs.~\cite{Li:2013zcr,Shen:2016tzq,Wu:2016ypc,Zhang:2018eeo,Liu:2019dqc,Xie:2019iwz,Wang:2020duv}. However, as discussed in Refs.~\cite{Hsiao:2019ait,Du:2019idk}, the ultraviolet divergences in the triangle loop diagrams integrals cancel out (for more details see Ref.~\cite{Achasov:2015uua}), thus we do not need to introduce these form factors in this work.

In Ref.~\cite{Hsiao:2019ait} the two-body scattering amplitude $t_{\pi\eta \to \pi \eta}$ is parameterized with the Breit-Wigner form. In this work, we describe the final state interaction between $\pi$ and $\eta$ as well as the interaction between $K$ and $\bar{K}$ with the chiral unitary approach. The scattering amplitudes $t_{\pi \eta \to\pi\eta}$ and $t_{K\bar{K}\to\pi\eta}$ can be obtained by solving the following Bethe-Salpeter equation
\begin{equation}
t=[1-V G]^{-1} V,
\end{equation}
where $G$ is the loop function of two mesons and $V$ is the transition potential. The loop function can be regularized by  either the dimensional regularization scheme or the  cutoff regularization scheme. In this work we employ the dimensional regularization scheme. The potential $V$ is a $2 \times 2$ matrix of coupled channels $K \bar{K}$ and $\pi\eta$. At the leading chiral order, the transition potential $V$ can be explicitly written as~\cite{Oller:1997ti,Xie:2014tma,Liang:2014ama,Liang:2015qva,Xie:2018rqv,Duan:2020vye}
\begin{align}
V_{K \bar{K} \to K \bar{K}}  = -\frac{1}{4 f^{2}} s, ~
V_{K \bar{K} \to \pi \eta}  = \frac{\sqrt{6}}{12 f^{2}}\left(3 s-\frac{8}{3} m_{K}^{2}-\frac{1}{3} m_{\pi}^{2}-m_{\eta}^{2}\right), ~
V_{\pi \eta \to \pi \eta}  = -\frac{1}{3 f^{2}} m_{\pi}^{2}, \nonumber
\end{align}
where we take the isospin multiplets as $K = (K^+,K^0)$, $\bar{K}= ( \bar{K^0},-K^-)$, and $\pi = (-\pi^+, \pi^0, \pi)$. Then, solving the Bethe-Salpeter equation with $\mu=1$ GeV, $a(\mu)_{\pi\eta}=-1.522$, $a(\mu)_{\bar{K}K}=-1.499$~\cite{Oller:2000fj}, we obtain a resonance with mass $m=983.2$ MeV  and width $\Gamma=105.6$ MeV. This can be associated with the $a_0(980)$ state.

With the so-obtained decay amplitudes, one can calculate the invariant mass distributions of $D^+_s\to \eta\pi^+\pi^0$ as a function of $m_{12}^2$ and $m_{23}^2$~\cite{Zyla:2020zbs}:
\begin{equation}
d\Gamma =  \frac{1}{(2 \pi)^{3}} \frac{|\mathcal{A}|^{2}}{32 m_{D_{s}}^{3}} d m_{12}^{2} d m_{23}^{2},
\end{equation}
where ${\cal A}$ is the total decay amplitude, which is ${\cal A} = {\cal A}^{\rm tree}_1 + {\cal A}_2^{\rho \eta} + {\cal A}_3^{K^* K}$.

One can easily obtain the single differential invariant mass distribution $d\Gamma/dm_{\pi\pi}$ and $d\Gamma/dm_{\pi \eta}$ by integrating over $m_{\pi \eta}$ and $m_{\pi \pi}$ with the limits of the Dalitz Plot, respectively.

\section{Numerical Results and Discussion}\label{sec:Results}

\begin{figure}[htbp]
  \centering
  \includegraphics[scale=0.4]{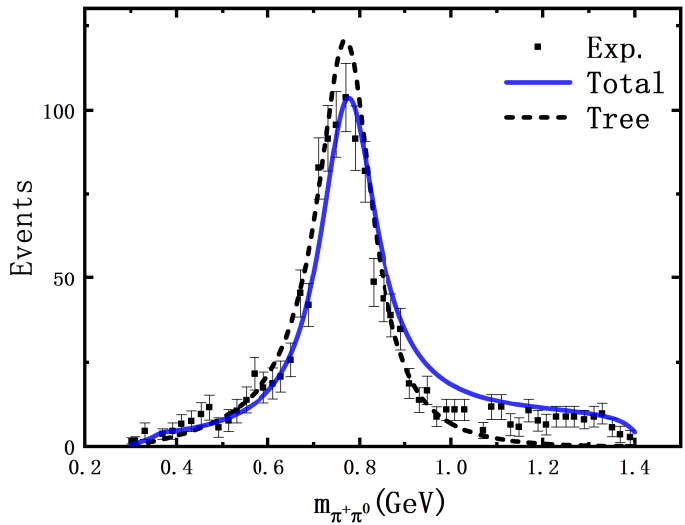}
  \caption{Invariant mass distribution of $\pi^+\pi^0$  for the $D_s^{+}\to \eta(\rho^+\to)\pi^0\pi^{+}$ decay, in comparison with the experimental data taken from Ref.~\cite{Ablikim:2019pit}.}\label{tree-pipi}
\end{figure}

We first show the theoretical results for the $\pi^+\pi^0$ invariant mass distribution in Fig.~\ref{tree-pipi} in comparison with the BESIII data~\cite{Ablikim:2019pit}. The solid curve stands for the total contributions from the tree diagram and the triangle diagrams, while the dashed curve stands for the contribution from only the tree diagram. The solid curve has been adjusted to the strength of the experimental data of BESIII at its peak~\cite{Ablikim:2019pit}.  Since we have considered the tree diagram contribution from the $\rho^+$ meson, one can see that the $\rho^+$ peak can be well reproduced. Furthermore, the high energy points for the $\pi^+ \pi^0$ invariant mass distributions can  also be well reproduced by including the contributions from the triangle diagrams. It is interesting to mention that the interference between the tree diagram and triangle diagrams is destructive below $m_{\pi^+\pi^0} = 0.8$ GeV, while above that energy point, the interference is constructive. Besides, with $a_1$, $a_2$, and $a_2'$ determined as specified above, we obtain an absolute branching ratio of $\mathcal{B}\left(D_{s}^{+} \to \pi^{+(0)}\left(a_{0}^{0(+)} \to\right) \pi^{0(+)} \eta\right)=1.45\%$. This is in nice agreement with the BESIII measurement.

\begin{figure}[htpb]
  \centering
\includegraphics[scale=0.35]{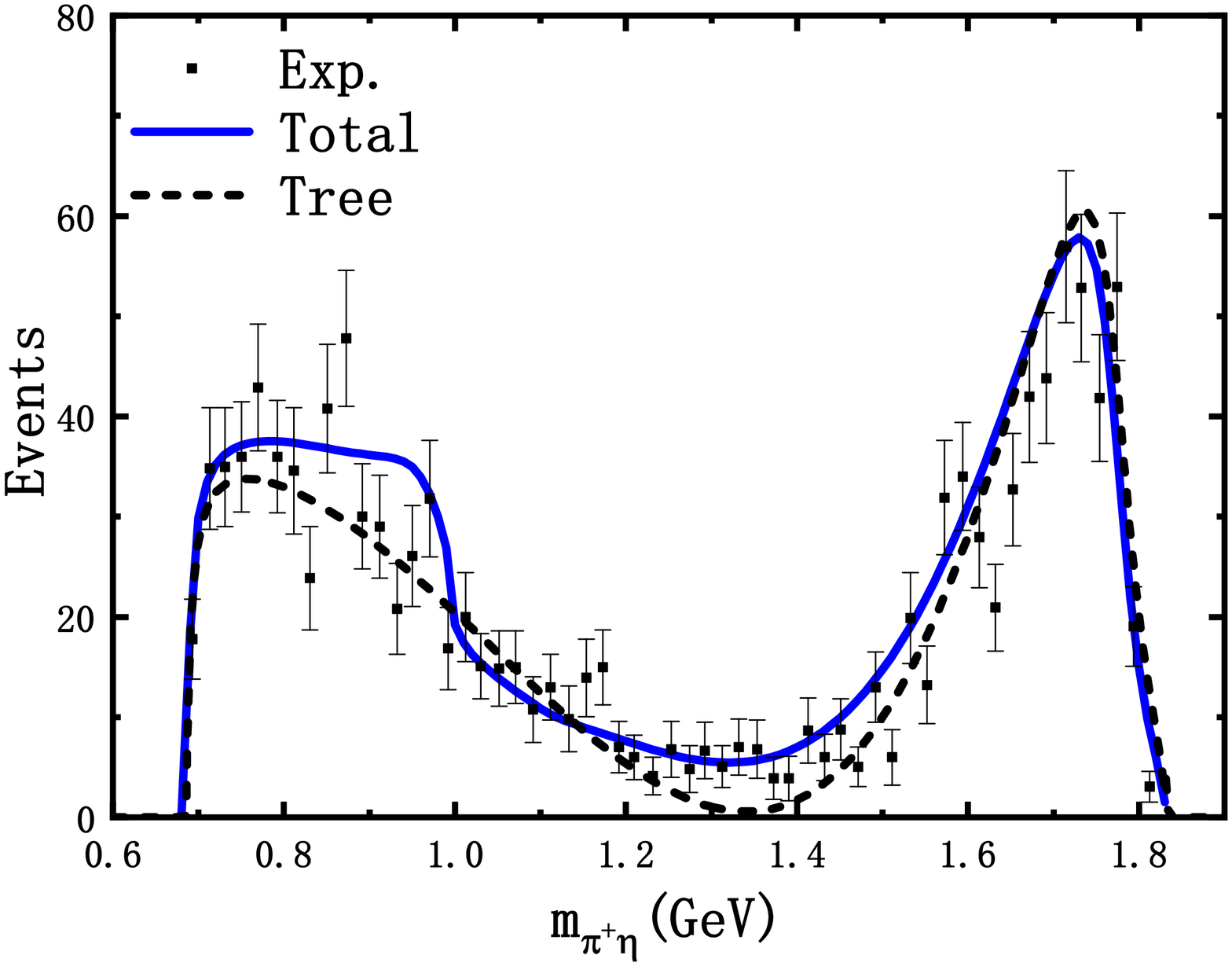}~
\includegraphics[scale=0.35]{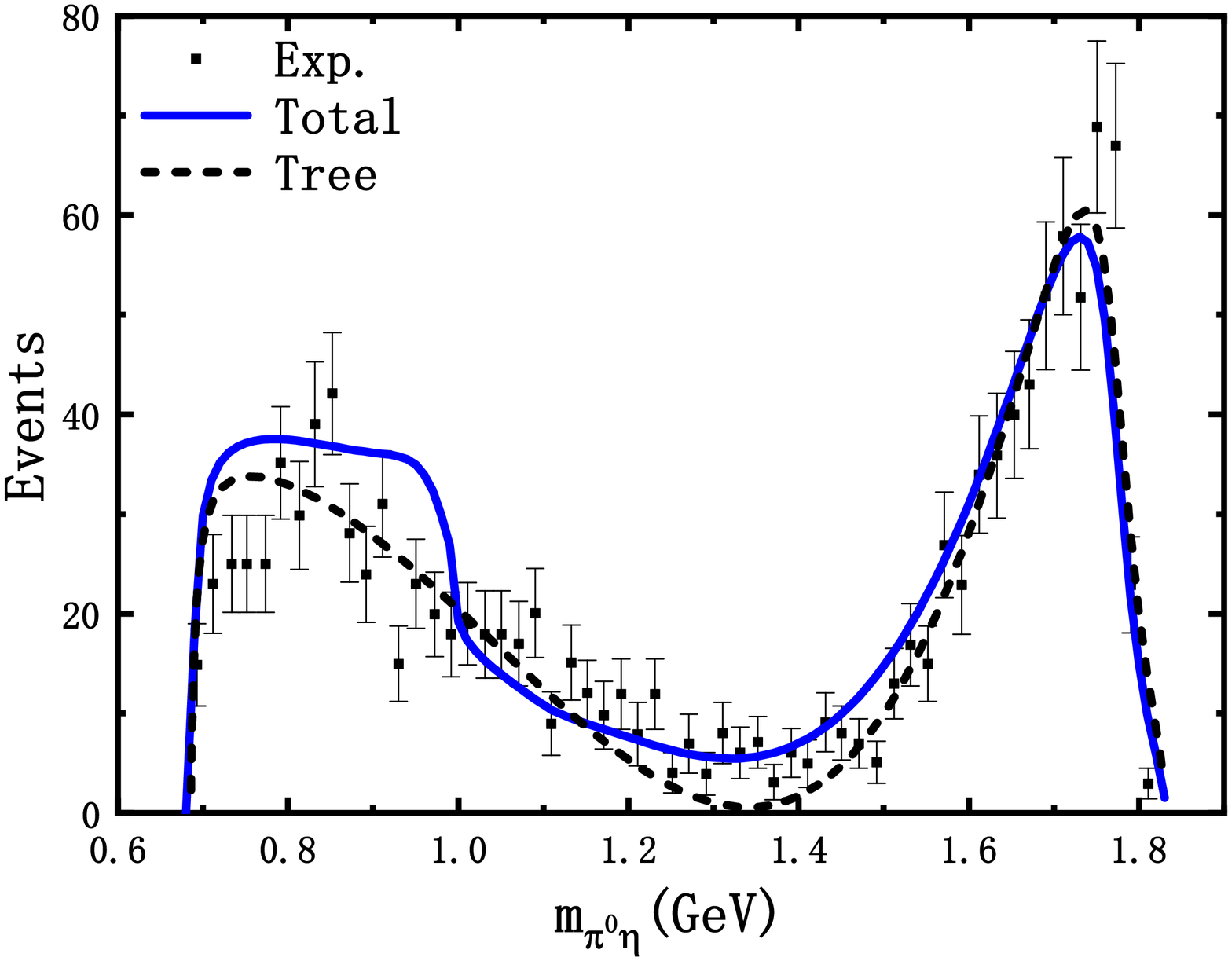}
  \caption{Invariant mass distributions of $\pi\eta$ for the $D_s^{+}\to \pi (a_0(980)\to)\pi^0\pi^{+}\eta$ decay, in comparison with the BESIII data~\cite{Ablikim:2019pit}.}\label{pi eta}
\end{figure}

\begin{figure}[htpb]
  \centering
  \includegraphics[scale=0.35]{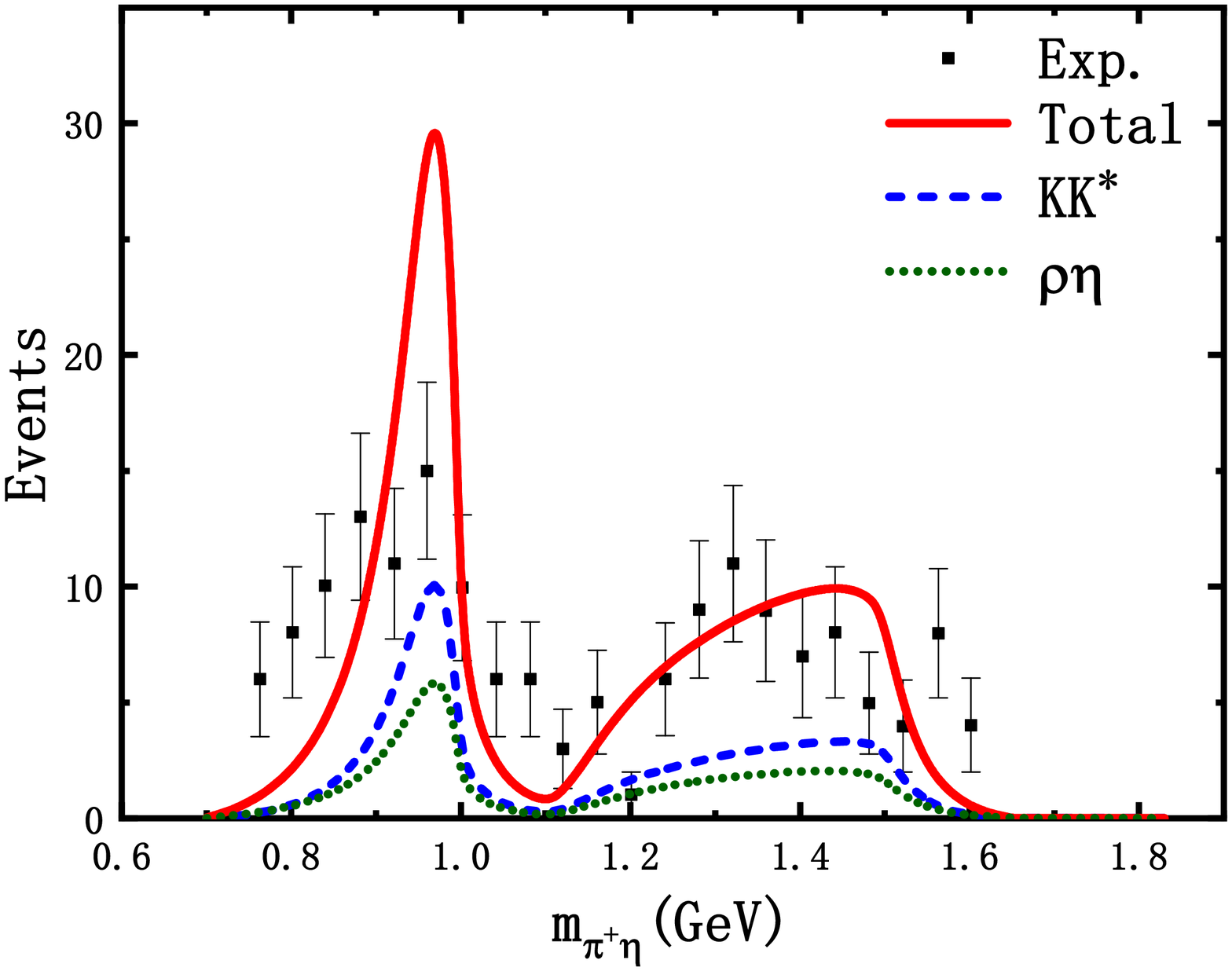}~
    \includegraphics[scale=0.35]{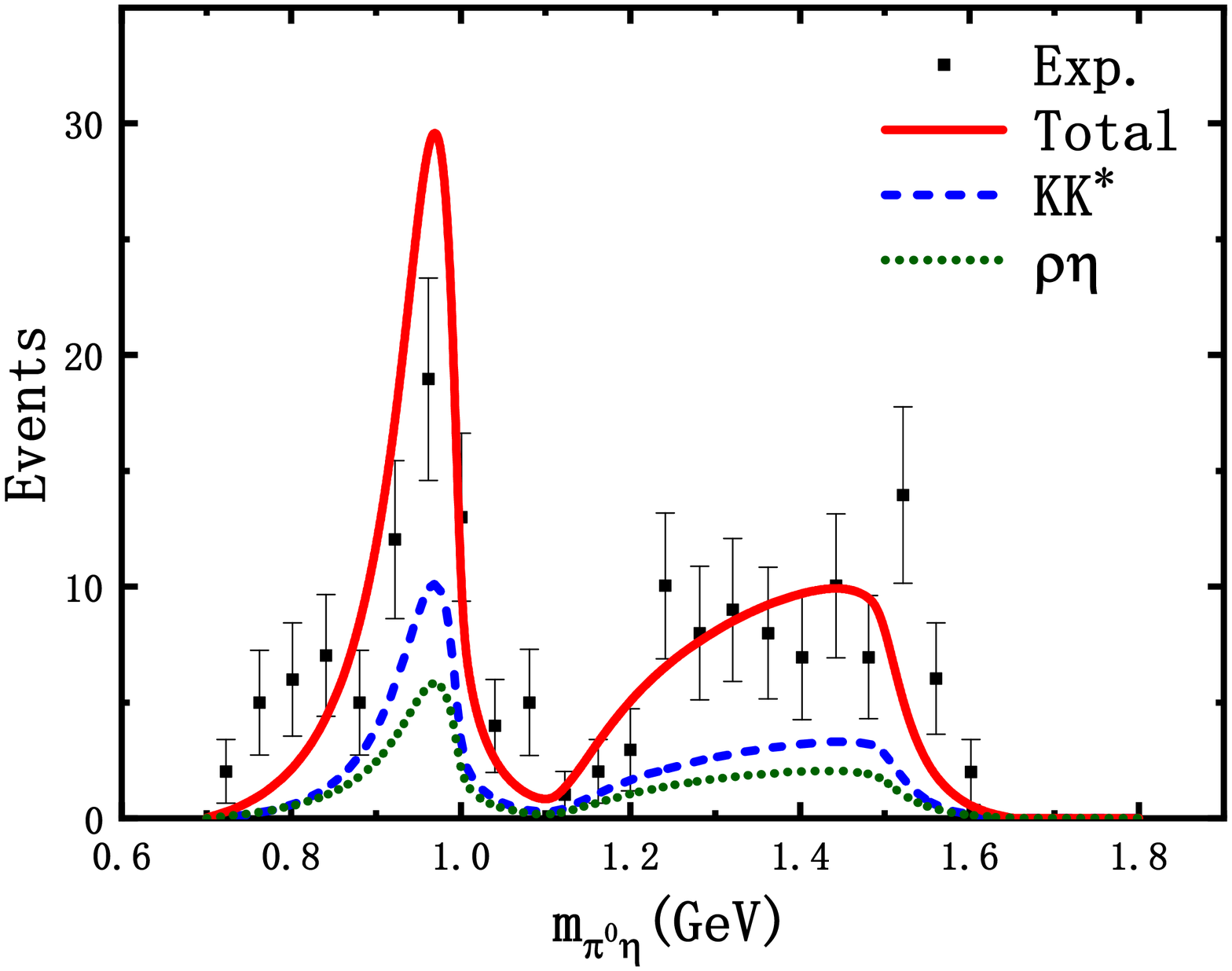}
  \caption{Invariant mass distributions of $\pi\eta$ with the cut of $m_{\pi^+\pi^0}> 1$ GeV for the decay $D_s^{+}\to \pi (a_0(980) \to)\pi^0\pi^{+}\eta$, in comparison with the BESIII data~\cite{Ablikim:2019pit}.}\label{pi eta cut}
\end{figure}

 In Fig.~\ref{pi eta} and Fig.~\ref{pi eta cut}, we show the $\pi^{0}\eta$ and $\pi^{+}\eta$ invariant mass distributions of the decay $D_{s}^{+}\to \pi^{+}\pi^{0}\eta$ without and with the cut of $m_{\pi^+\pi^0}>1$ GeV, respectively. From Fig.~\ref{pi eta}, one can see that the contribution from the tree diagram is predominant. The theoretical results can describe the experimental data rather well, particularly the shoulder around $m_{\pi \eta} \sim 0.8$ GeV. In addition, our theoretical results do not show a pronounced asymmetric peak around $m_{\pi \eta} \sim 0.9$ GeV as in Ref.~\cite{Hsiao:2019ait} (see Fig. 4 of that reference).

In Fig.~\ref{pi eta cut}, the dashed curves represent the contributions from the $\rho \eta$ triangle diagrams shown in Fig.~\ref{triangle}, while the blue-dashed curves represent the contributions from the $\bar{K}K^*$ triangle diagrams shown in Fig.~\ref{triangle KK}, and the red-solid curves stand for the sum of the two contributions. From Fig.~\ref{pi eta cut}, one can see that after the $m_{\pi^+\pi^0}>1$ GeV cut, the $a_0(980)$ signal is well reproduced, where it is dynamically generated from the $\pi\eta$ and $K\bar{K}$ coupled channel interactions. However, our results at the $a_0(980)$ peak position are somehow larger than the experimental data, especially for the case of $a_0(980)^+$. It should be noted that in  Ref.~\cite{Molina:2019udw}, the $\pi K \bar{K}$ final states were produced at the first step with the internal $W$-emission mechanism, and then the final state interaction of $K\bar{K}$ produces $a_0(980)$, which then decays to $\pi \eta$. Clearly, Ref.~\cite{Molina:2019udw} and the present work share the same mechanism for the final state interactions. As a result, both can describe the $\pi\eta$ line shapes, but the present work also determines the global strength of the $D_s$ decay. In principle, both weak mechanisms may play a role. However, a quantitative consideration of the mechanism of Ref.~\cite{Molina:2019udw} inevitably introduces additional free parameters for the weak interaction, which cannot yet be determined. Hence, we will leave such a study to a future work when more precise experimental data become available.

It is worthwhile mentioning that in our framework that the $K\bar{K}^*$ contribution is larger than the $\rho\eta$ contribution, while the former was not considered in Ref.~\cite{Hsiao:2019ait}, where the $\rho \eta'$ channel plays an important role and a large coupling for $a_0(980)$ to $\pi \eta'$  is used. However, both in the unitary chiral approach and from the experimental information, it is known that the $\pi\eta'$ coupling to the $a_0(980)$ resonance is small.

\begin{figure}[htpb]
  \centering
\includegraphics[scale=0.45]{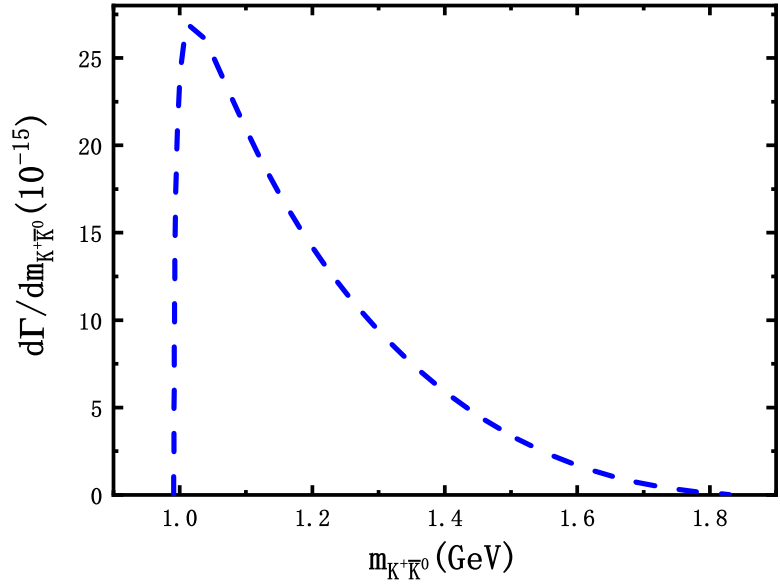}
  \caption{Invariant $K^+ \bar{K}^0$ mass distribution for the $D_s^+ \to \pi^0 K^+ \bar{K}^0$ decay.}\label{KKbar}
\end{figure}

In addition, we study the $a_0(980)$ state in the $K\bar{K}$ channel from the  $D^+_s \to \pi K \bar{K}$ decay by including the contributions from the triangle diagrams, which can be easily obtained with the replacement of the $\pi \eta$ final state by $K\bar{K}$ in Figs.~\ref{triangle} and \ref{triangle KK}. The resulting predictions for the $K^+\bar{K}^0$ invariant mass distributions are shown in Fig.~\ref{KKbar}, which can serve as a highly non-trivial check of the mechanism proposed in this work.

\section{Summary}

We studied the $D^+_s\to \pi^+\pi^0\eta$ decay recently analysed by the BESIII Collaboration, where the $D^+_s \to \pi^0 a_0(980)^+$ and $D^+_s \to \pi^+ a_0(980)^0$ decay modes are claimed as the $W$-annihilation dominant processes observed for the first time, and their branching fractions, however, are one order of magnitude larger than those of known $W$-annihilation decays. Inspired by Ref.~\cite{Hsiao:2019ait}, we proposed that the anomalously large branching ratios of these decay modes can be understood via triangle diagrams. At first, the $D^+_s$ meson decays weakly into either $\rho^+\eta$ or $K\bar{K}^*/K^*\bar{K}$. The vector mesons then decay into a pair of pseudoscalar mesons, $\pi\pi$ or $K\pi$. One of them interacts with the pseudoscalar meson from the weak decay of $D^+_s$, and generates dynamically the $a_0(980)$ state. With the weak decay couplings determined by fitting to the experimental branching fractions, our method predicted both the absolute branching ratio of $D_s\to a_0\pi$ and the $\pi\eta$ invariant mass distributions, which are in nice agreement with the BESIII data. For the $D^+_s \to \pi^+ \pi^0 \eta$ decay, the contribution from the tree diagram as shown in Fig.~\ref{tree} is the most dominant. After the cut of $M_{\pi^+\pi^0} >1$ GeV, the contributions of the triangle diagrams are crucial to produce the $a_0(980)$ resonance. In addition, we predicted the $K\bar{K}$ invariant mass distributions of the $D_s\to\pi K\bar{K}$ decay, which can be checked by future experimental measurements.

Furthermore, the present work provides a way to estimate the effective weak coupling appearing in the $D_s\to K^*\bar{K}\to K\bar{K}\pi$ vertex. The same mechanism might also be relevant to those of similar processes, such as the $D^+\to \pi^+ K\bar{K}$ and $D^+\to \pi^+ \pi^0 \eta$ reactions~\cite{Duan:2020vye,Ikeno:2021kzf}.

\label{sec:Summary}

\section*{Acknowledgments}
This work is partly supported by the National Natural Science Foundation of China under Grants No. 11735003, No. 11975041, No. 11775148, No. 12075288, and No. 11961141004. It is also supported by the Youth Innovation Promotion Association CAS (2016367).

{}

\end{document}